\input epsf
\documentstyle[12pt]{article}
\begin{document}
\begin{titlepage}
\rightline{Preprint YERPHI-1532(5)-99} 
\vspace{2cm}
\begin{center}
\large{\bf{Upper Bound on the Lightest Higgs Mass in Supersymmetric
Theories}} \\
\vspace{1cm}
\large{G. K. Yeghiyan} \\
{\em Yerevan Physics Institute, Yerevan, Armenia}\\
{\em e-mail: egagik@jerewan1.yerphi.am}
\end{center}
\vspace{1cm}
\centerline{\bf{Abstract}}
\vspace{0.1cm}
The problem of the lightest Higgs boson mass in the next-to-minimal
supersymmetric standard model (NMSSM) is investigated. Assuming the
validity of the perturbation theory up to
unification
scales and using the recent experimental results for the top quark mass, the
restrictions on the NMSSM coupling constants are obtained. These
restrictions are used to make the predictions for the lightest Higgs
mass, which are compared to those of the minimal supersymmetric standard
model (MSSM).
\end{titlepage}
1. The aim of this letter is to investigate the lightest Higgs boson mass
problem in the next-to-minimal supersymmetric standard model (NMSSM). This
model contains an additional Higgs singlet, as compared to the minimal
supersymmetric standard model (MSSM). The cause of consideration of the
NMSSM is connected in particular with desire to avoid an explicit mass
term for the Higgs doublets in a superpotential \cite{1}. There is also an
interesting possibility to realize the scenario with spontaneous CP-violation
in this model \cite{2}-\cite{6}. It is known that in the minimal
supersymmetric standard model such a scenario  \cite{7}
contradicts with existing
experimental constraints on the MSSM neutral Higgs masses  \cite{28,8},
while in the NMSSM this contradiction
can be avoided in different ways \cite{5,6}. The experimental
constraints on Higgs
masses play important role also for the case of the absence of
spontaneous CP violation. The recent LEP experiments give the lower bound
on the lightest MSSM
Higgs mass $m_h > 75$GeV  \cite{28}. Depending on the Higgs vacuum
expectation values (vev's) ratio
$\tan{\beta}$, this bound can be much stronger. Thus, one derives \cite{8}
$m_h > 88$GeV for $\tan{\beta} \leq 2$.  This makes the MSSM close to being
excluded, when the low $\tan{\beta}$ scenario is considered.
It is interesting to analyze how the situation is looked in the NMSSM.           \\
\\
2. The next-to-minimal supersymmetric standard model had been widely studied
in literature \cite{9}-\cite{17}. It was noted already in ref. \cite{9} that
the upper bound on the lightest NMSSM Higgs particle mass
can be different from the
one in the MSSM. On the other hand, the NMSSM Higgs sector is described (even
at tree level) by large enough number of unknown parameters, which makes the
analysis of Higgs masses very difficult. Today only the case, when the
supersymmetry breaking
parameters of the theory are related by universality conditions, has been
investigated in details \cite{12,14,15}.  Requiring for the physical
minimum of the scalar potential to be the global one and taking into
account the experimental
bounds on the NMSSM particles masses, it was found that the Higgs
singlet sector is decoupled and -  due to smallness of the relevant coupling
constants - the predictions for the detectable
(nonsinglet) Higgs bosons are almost the same as those of the MSSM.
However,
as it was recently pointed out \cite{16}, the NMSSM parameters space, where
the scenario with the universal supersymmetry breaking can be realized, is
strongly restricted, if it is required for the physical minimum of the scalar
potential to be the global one. One can avoid such a strong constraints on
parameters of theory either assuming that the physical minimum of the
potential is a local one, or relaxing the universality conditions at
high energy scales. \\
In this letter no universality conditions on the supersymmetry breaking
parameters are assumed. In this case either only some special
regions of parameters space of theory are investigated or an upper bound on
the
mass of the lightest Higgs boson, as a function of some parameter(s) of
theory, is only found. Here the latter strategy is
used: an upper bound on the lightest Higgs mass, as a function of
$\tan{\beta}$, is obtained. The only restrictions on parameters of theory,
which
are taken into account, are constraints on the coupling constants,
coming from the assumption of the absence of new physics between the
supersymmetry breaking scale ($M_{SUSY}$) and unification scales ($M_G$), as
well as from the recent experimental results for the top quark mass.
The effects, connected with new experimental results for the top quark mass,
play important role, when being taken into account appropriately. These
effects together with two-loop order corrections to the lightest Higgs mass
lead to the predictions for this particle mass, which are
different by about (20-25)GeV from those, reported in literature
previously. The derived
upper bound on the lightest Higgs mass is compared to the one
of the MSSM.  \\
\\
3. The NMSSM superpotential is obtained from the MSSM one by making  in the
latter one the  following replacement:
\begin{equation}
\mu \bar{H}_1 \bar{H}_2 \to \lambda \bar{N} \bar{H}_1 \bar{H}_2 +
\frac{\kappa}{3} \bar{N}^3  
\end{equation}
(here $\bar{H}_1$ and $\bar{H}_2$ are the Higgs doublets and $\bar{N}$ is the
Higgs singlet superfields with scalar components $H_1$, $H_2$, N).
Thus, one avoids an explicit mass term in the superpotential - such a term
was needed, to
provide for spontaneous breaking of the $SU(2)_L \times U(1)_Y$ symmetry.
Instead one introduces the additional complex degree of freedom, connected
with the $SU(2)_L \times U(1)_Y$ Higgs singlet N, as well as two Yukawa-type
coupling constants, $\lambda$ and
$\kappa$. These couplings enter the Higgs bosons mass matrices in nontrivial
way and to find the restrictions on the lightest Higgs mass, one has to
obtain some
constraints on $\lambda$ and $\kappa$ and/or correlations between them.
Such constraints can be obtained from the NMSSM renormalization
group equations (RGE's) analysis. Such an analysis in the standard model and 
its extensions has been performed in ref's  \cite{18}-\cite{27} - there
various approaches had been developed. \\
In this letter it is assumed that all the (gauge and Yukawa-type)
couplings are
small between the electroweak breaking scale $\sim 100$GeV (this scale is
identified also with Z boson mass $M_Z$ or with the top quark mass $m_t$)
and unification scales ($M_G \sim (10^{16}-10^{18})$GeV), so that the
perturbation theory is applied. This condition leads to restrictions on
coupling
constants of theory. As a result, one obtains some bounds on the lightest
Higgs boson mass. \\
During the renormalization group equations analysis I neglect the
supersymmetric particles mass thresholds effect, assuming that
supersymmetric
RGE's are valid from the electroweak breaking scale up to unification scales.
I perform the analysis at one-loop level - as the numerical analysis shows,
the two-loop corrections do not exceed few percents, i.e. are of the
same order, as the inaccuracy of approach, connected with the uncertainty of
choice of the cut-off scale (whether it is chosen $\sim 10^{16}$GeV or, say,
$\sim 10^{18}$GeV). \\
\\
4. Let us discuss briefly the problem of the lightest Higgs boson mass
in the MSSM and its simplest extension with the additional Higgs
singlet. In the
minimal supersymmetric standard  model \cite{31,32}
there are five physical Higgs states:
two CP-even, one CP-odd and one complex charged Higgs states. At tree level
these particles masses are described by only two unknown parameters: the
CP-odd Higgs mass $m_A$ and the Higgs doublets vev's ratio
$\tan{\beta}=v_2/v_1$. As a result, Higgs boson masses and the restrictions
on them are given by compact expressions. In particular, for the lightest
(CP-even)
Higgs mass it is obtained that $m_h < M_Z |\cos{2\beta}|<M_Z$. Notice however
that this relation is violated, when radiative corrections to the Higgs
potential are taken into account. Usually one takes into account only the
radiative corrections, connected with the top and bottom quarks and squarks
loops - this is because of largeness of the top and (for large $\tan{\beta}$)
bottom Yukawa couplings, as compared to other coupling constants. When
being taken into account at one-loop level \cite{33}-\cite{37}, these
corrections can increase the lightest Higgs mass about (30-60)GeV,
depending on parameters space of theory. Recently the two-loop order
corrections
to the lightest Higgs mass had been also calculated \cite{38}-\cite{40}.
It was found that these corrections are important too: they can
lower $m_h$ more than 10GeV.  \\
The above-mentioned radiative corrections to the lightest Higgs mass depend
on the following set of parameters: the top mass $m_t$, the stop masses
$m_{\tilde{t}_1}$, $m_{\tilde{t}_2}$, the sum of Higgs
doublets vev's squared
$\eta^2=v_1^2+v_2^2=(174GeV)^2$, the Higgs vev's ratio $\tan{\beta}$,
the QCD coupling constant
$\alpha_3=g_3^2/(4 \pi)$, as well as (via the stops mixing effects) on the
stop trilinear supersymmetry breaking parameter $A_t$ and the explicit mass
parameter $\mu$. In this paper it is assumed that
$m_{\tilde{t}_1}^2 - m_{\tilde{t}_2}^2 \ll m_{\tilde{t}_1}^2 +
m_{\tilde{t}_2}^2$
(then stops masses are identified usually with the supersymmetry breaking
scale). In this approximation one obtains at two-loop level \cite{39,40}
\begin{eqnarray}
\nonumber
m_h^2 < M_Z^2 \cos^2{2\beta} \left(1-\frac{3}{8 \pi^2} \frac{m_t^2}{\eta^2}
\log{\frac{M_{SUSY}^2}{m_t^2}}\right) + \frac{3}{4 \pi^2}
\frac{m_t^4}{\eta^2}\Biggl[\frac{1}{2} X_t + \\ +
\log{\frac{M_{SUSY}^2}{m_t^2}} + \frac{1}{16 \pi^2} \left(\frac{3}{2}
\frac{m_t^2}{\eta^2} - 32 \pi \alpha_3 \right) \left(X_t             
\log{\frac{M_{SUSY}^2}{m_t^2}} + \log^2 {\frac{M_{SUSY}^2}{m_t^2}} \right)
\Biggr]
\end{eqnarray}
where
\begin{equation}
X_t = \frac{2(A_t - \mu \cot{\beta})^2}{M^2_{SUSY}} \left(1 -
\frac{(A_t - \mu \cot{\beta})^2}{12 M^2_{SUSY}}\right)   
\end{equation}
It is easy to see that $X_t \leq 6$ - the maximum value of $X_t$ is obtained,
when $A_t - \mu \cot{\beta} = \sqrt{6} M_{SUSY}$. In this paper two values
of $X_t$ are considered: $X_t=0$ - the case of so-called no-mixing, and
$X_t=6$ - the case of so-called maximal mixing. \\
In the next-to-minimal supersymmetric standard model there are additional
one CP-even and one CP-odd Higgs
degrees of freedom, as compared to the MSSM. At tree level the $3 \times 3$
symmetric mass matrix
for CP-even Higgs fields $\Phi_1$, $\Phi_2$ and $N_1$ \cite{9}-\cite{11},
\cite{17} is given by formula (A1) of Appendix. Usually one takes
$\Phi_1=\frac{1}{\sqrt{2}}(Re(H_1^0) - v_1)$, $\Phi_2=\frac{1}{\sqrt{2}}
(Re(H_2^0)-v_2)$ and $N_1=\frac{1}{\sqrt{2}}(Re(N)-v_3)$, where $v_3$ is the
singlet vev.  In this paper a little bit
different representation is used, namely,
\begin{eqnarray}
\nonumber
\Phi_1&=&\frac{1}{\sqrt{2}}(Re(H_1^0)\cos{\beta} + Re(H_2^0) \sin{\beta}-
\eta)\\
\Phi_2&=&\frac{1}{\sqrt{2}}(-Re(H_1^0)\sin{\beta} + Re(H_2^0) \cos{\beta}) \\
\nonumber
N_1&=&\frac{1}{\sqrt{2}}(Re(N)-v_3)   
\end{eqnarray}
Such a choice of the representation of CP-even Higgs fields makes the matrix
(A1) more transparent for the qualitative analysis. \\
At tree level the upper bound on the lightest (CP-even) Higgs boson mass is
the following:
\begin{equation}
m_h^2 < M_Z^2 \cos^2{2\beta} + \lambda^2 \eta^2 \sin^2{2\beta}  
\end{equation}
where the right hand side of eq. (5) is the maximum value of the lowest
eigenvalue of $2 \times 2$ upper block ($M_{S_{ij}}^2, i,j=1,2$) of the
matrix (A1) \cite{10,11}. In other words, the upper bound (5) is saturated,
when the singlet sector is decoupled\footnote{The necessary condition for
this to occur is $M_{S_{13}}^2 \to 0$. If even
$M_{S_{22}}^2,M_{S_{33}}^2 \gg M_{S_{11}}^2 \sim M_Z^2$ , one derives
$m_h^2 = M_{S_{11}}^2 - (M_{S_{13}}^2)^2/M_{S_{33}}^2$, so that the upper
bound (5) is saturated, when $M_{S_{13}}^2 \to 0$. This occurs if either
$\lambda \to 0$, or $v_3 \to 0$ (this case is disfavoured by the
experimental bounds on the
lightest chargino mass) or in the case of fine-tuning of the relevant NMSSM
mass parameters. In other words, the upper bound (5) is
saturated only in narrow region of the NMSSM parameters space. It is worth
to stress also that this is the upper bound on the lightest {\em detectable}
(nonsinglet) Higgs boson.}
and the detectable Higgs bosons masses are
described by the $2 \times 2$ MSSM upper block of the matrix (A1).  Like in
the MSSM, there is only one (nonsinglet) Higgs boson with the mass of the
order of electroweak breaking scale, when the upper bound on $m_h$ is
saturated\footnote{Generally speaking, this rule is violated for
$\tan{\beta} \gg 1$: then the upper bound on $m_h$ can be saturated also
when more than one Higgs bosons with the masses $\sim M_Z$ exist. However
these values of $\tan{\beta}$ are out of the interest:
as it is easy to understand, the difference between the models predictions
occurs due to the coupling $\lambda$, which enters the above-mentioned
$2 \times 2$ block with the factor $\sin{2 \beta}$.}. \\
Even in the region of parameters space, where the singlet sector is
decoupled, there is a difference between the radiative corrections to
the lightest Higgs mass in the MSSM and those in the NMSSM. Such
a difference arises due to the loops, connected with the term
$(\lambda^2 - g_2^2/2) |H_1 H_2|^2$ of the Higgs potential \cite{10}. The
contribution of these loops is
proportional to $\lambda^4$, $\lambda^2 g_2^2$, $\lambda^2 g_1^2$,
$\lambda^2 h_t^2$, etc. ($g_2$ and $g_1$ are respectively $SU(2)_L$ and
$U(1)_Y$ gauge coupling constants, $h_t$ is the top quark Yukawa coupling).
\\
In the next section it is obtained that $\lambda$ is of the order of weak
coupling constant $g_2$ or smaller. This allows one to neglect
all aforementioned corrections, except of
those, proportional to $\lambda^2 h_t^2$. These latter corrections arise due
to the top quark and squark loops. The contribution of these loops is
generalized in the case of the NMSSM in the following way. \\
As it was mentioned above, both in the MSSM and in the NMSSM the upper bound
on the lightest Higgs mass is saturated, when there is only one light
(detectable) Higgs boson with the mass of the order of electroweak breaking
scale. Then at tree level one may write
$m_h^2 < m_{h_{max}}^2= 4 \lambda_{SM} \eta^2$, where\footnote{The choice of
the coefficient in front of $\lambda_{SM}$ depends on the choice of
the corresponding coefficient of the standard model Higgs potential.
Here the same
notation for $\lambda_{SM}$, as in ref. \cite{26}, is used.}
$\lambda_{SM}=M_Z^2\cos^2{2\beta}/(4 \eta^2)$ and
$\lambda_{SM}=M_Z^2\cos^2{2\beta}/(4 \eta^2) + \lambda^2 \sin^2{2\beta}/4$
for the MSSM and the NMSSM respectively. As for the radiative corrections,
they can be subdivided in following three parts:
\begin{quote}
a) the corrections, coming from the one-loop renormalization of
$\lambda_{SM}$, \\
b) one-loop corrections, connected with the stops mixing effect, \\
c) next-to-leading order (two-loop order) corrections.
\end{quote}
The corrections a)  can be read off from the standard model RGE's \cite{26}:
one obtains that
\begin{displaymath}
\lambda_{SM} \to \lambda_{SM} \left(1 - \frac{3}{8\pi^2} h_{t_{SM}}^2
\log{\frac{M_{SUSY}^2}{m_t^2}}\right) + \frac{3}{16 \pi^2} h_{t_{SM}}^4
\log{\frac{M_{SUSY}^2}{m_t^2}}
\end{displaymath}
where $h_{t_{SM}}=m_t/\eta$. The corrections b) and c) depend only on the
top and stop sectors of
Lagrangians of the models. These sectors (and hence the corrections b) and
c)) are almost the same in both of models. The difference is only that the
mass parameter $\mu$ is replaced by the product $\lambda N$ (or vice
versa). Respectively,
in equation (3) one has to make the replacement $\mu \to \lambda v_3$. \\
Using the above-mentioned analogy between the models, one obtains the
following bound on the lightest NMSSM Higgs particle:
\begin{eqnarray}
\nonumber
m_h^2 < (M_Z^2 \cos^2{2\beta} + \lambda^2 \eta^2 \sin^2{2\beta}) 
\left(1-\frac{3}{8 \pi^2} \frac{m_t^2}{\eta^2}
\log{\frac{M_{SUSY}^2}{m_t^2}}\right) 
+ \frac{3}{4 \pi^2}
\frac{m_t^4}{\eta^2}\Biggl[\frac{1}{2} X_t + \\ +
\log{\frac{M_{SUSY}^2}{m_t^2}} + \frac{1}{16 \pi^2} \left(\frac{3}{2}
\frac{m_t^2}{\eta^2} - 32 \pi \alpha_3 \right) \left(X_t             
\log{\frac{M_{SUSY}^2}{m_t^2}} + \log^2 {\frac{M_{SUSY}^2}{m_t^2}} \right)
\Biggr]
\end{eqnarray}
One can see from comparison of (6) to (2) that the
difference between the models predictions can occur due to the term,
proportional to
$\lambda^2 \eta^2 \sin^2{2\beta}$. This term is important for low values
of $\tan{\beta}$. Therefore I restrict myself by consideration of
$\tan{\beta} \leq 6$. To estimate
the difference between the models predictions, one must find some
constraints on the coupling constant $\lambda$. As it was mentioned above,
such constraints are derived from the renormalization group equations
analysis. \\
One assumes often (see e.g. \cite{17}) that $|\lambda| < 0.87$-
the infrared (IR) quasi-fixed
point, derived in the limit $h_t=0$.
Such an assumption
is not quite correct: the restrictions on $\lambda$ become stronger, when
$h_t \sim 1$ \cite{10,11}. As it is shown in the next section, the
restrictions on $\lambda$ for the experimentally allowed range of the top
mass are about 20$\%$ different from the bound, reported above. \\
\\
5. For the low $\tan{\beta}$ scenario the RGE's for $SU(3)_c$, $SU(2)_L$,
$U(1)_Y$ couplings, $g_3$, $g_2$ and $g_1$ respectively, and Yukawa-type
couplings $h_t$, $\lambda$ and $\kappa$ are the following \cite{27,41}:
\begin{eqnarray}
\nonumber
 16\pi^{2}\frac{d\kappa}{dt} & = & 6\kappa (\kappa^{2} +
\lambda^{2}) \\
16\pi^{2}\frac{d\lambda}{dt} & = & \lambda (2 \kappa^{2} +
4\lambda^{2} + 3 h_{t}^{2} - \frac{3}{5} g_{1}^{2} -3g_{2}^{2})  \\
\nonumber
16\pi^{2}\frac{dh_{t}}{dt} & = & h_{t}(6h_{t}^{2} +
\lambda^{2} -
\frac{13}{15}g_{1}^{2} - 3g_{2}^{2} - \frac{16}{3}g_{3}^{2})  \\
\nonumber
16\pi^{2}\frac{dg_{i}}{dt} & = & -c_{i}g_{i}^{3}
\end{eqnarray}
where $c_1=-33/5$, $c_2=-1$, $c_3=3$ and $t=1/2\ \log(Q^2/M_Z^2)$ (as it was
mentioned above, I neglect the supersymmetric particles mass thresholds
effect,
considering these RGE's valid from the electroweak breaking scale up to
unification scales).
The behaviour of gauge couplings is determined by their experimental values
at the electroweak breaking scale \cite{42}:
$g_1(m_t) \approx 0.46$,
$g_2(m_t) \approx 0.65$, $g_3(m_t) \approx 1.22$. Some restrictions on the
top quark
Yukawa coupling is found from the experimental constraints on the top mass.
According the recent experimental data \cite{42},
$m_t^{pole}=(173.8 \pm 5.2)$GeV. One has to transform the pole top mass into
the on-shell top mass
\begin{equation}
m_t \equiv m_t(m_t)=h_t(m_t) \eta \sin{\beta} 
\end{equation}
to make the predictions for $h_t$. This transformation is done, using
the
well-known relation between the pole and the on-shell top masses \cite{43};
\begin{equation}
m_t(m_t)=\frac{m_t^{pole}}{1+ \frac{g_3^2}{3 \pi^2}}=(165 \pm 5)GeV 
\end{equation}
where in eq. (9) only the leading order QCD gluon corrections are taken into
account. Higher order QCD corrections as well as the stop/gluino loops
corrections \cite{44,45}
modify the on-shell top mass by about 2GeV - they can also
cancel each other.  Here I neglect these corrections: to take into account
the stop/gluino loops effects, one
has to determine both the stops masses and the stops mixing angle.
This deviates from the framework of this letter.\\
As it follows from equations (8), (9), $0.92/\sin{\beta} \leq h_t(m_t) \leq
max(h_{t_{max}},0.98/\sin{\beta})$, where $h_{t_{max}}$ is the triviality
bound on $h_t(m_t)$. This bound, as well as bounds on $\lambda$ and $\kappa$
($\lambda_{max}$ and $\kappa_{max}$)
are found from the analysis of RGE's for these coupling constants. \\
The procedure is the following. One evolutes (numerically)
$\lambda$, $\kappa$, $h_t$ and
$g_1$, $g_2$, $g_3$ according their renormalization group equations
from the electroweak breaking scale up to the GUT scale
$\sim 10^{16}$GeV. It is required at whole considered energy range for the
couplings to be small enough to perturbation theory being applied. More
precisely, the following condition on $\lambda$, $\kappa$ and $h_t$ must be
satisfied:
\begin{equation}
\lambda^2(Q^2) < 4 \pi,\ \kappa^2(Q^2) < 4 \pi,\ h_t^2(Q^2) < 4\pi 
\end{equation}
The conditions (10) put some constraints on coupling constants at the
electroweak breaking scale. Thus, one
obtains $h_t(m_t) < 1.15$ and $|\kappa(m_t)| < 0.65$. The reported upper
bound on $h_t(m_t)$ is valid also for the MSSM: this bound is obtained in the limit
$\lambda =0$, where the RGE for $h_t$ is reduced to the MSSM one.
Using equation (8),
one can also transform (for fixed value of the top mass) the restriction on
$h_t(m_t)$ to the lower bound on $\tan{\beta}$. Clearly, this lower bound
is the same both for the MSSM and for the NMSSM. \\
The restrictions on
$\lambda$, as functions of $\tan{\beta}$, are presented in Fig,~1.
\begin{figure}
\caption{Upper bound on $|\lambda(m_t)|$ ($\lambda_{max}$) as a function of
$\tan{\beta}$ 
a) for $\kappa(m_t)=0$ and $m_t=160$GeV (line 1), $m_t=165$GeV (solid line)
and $m_t=170$GeV (line 3),
b) for $m_t=160$GeV and $|\kappa(m_t)|=0;0.3;0.4;0.5;0.6$ (lines 1,2,3,4,5
respectively)}
\epsfxsize = 15cm
\epsfysize = 9cm
\mbox{\hskip 0.4in \epsfbox{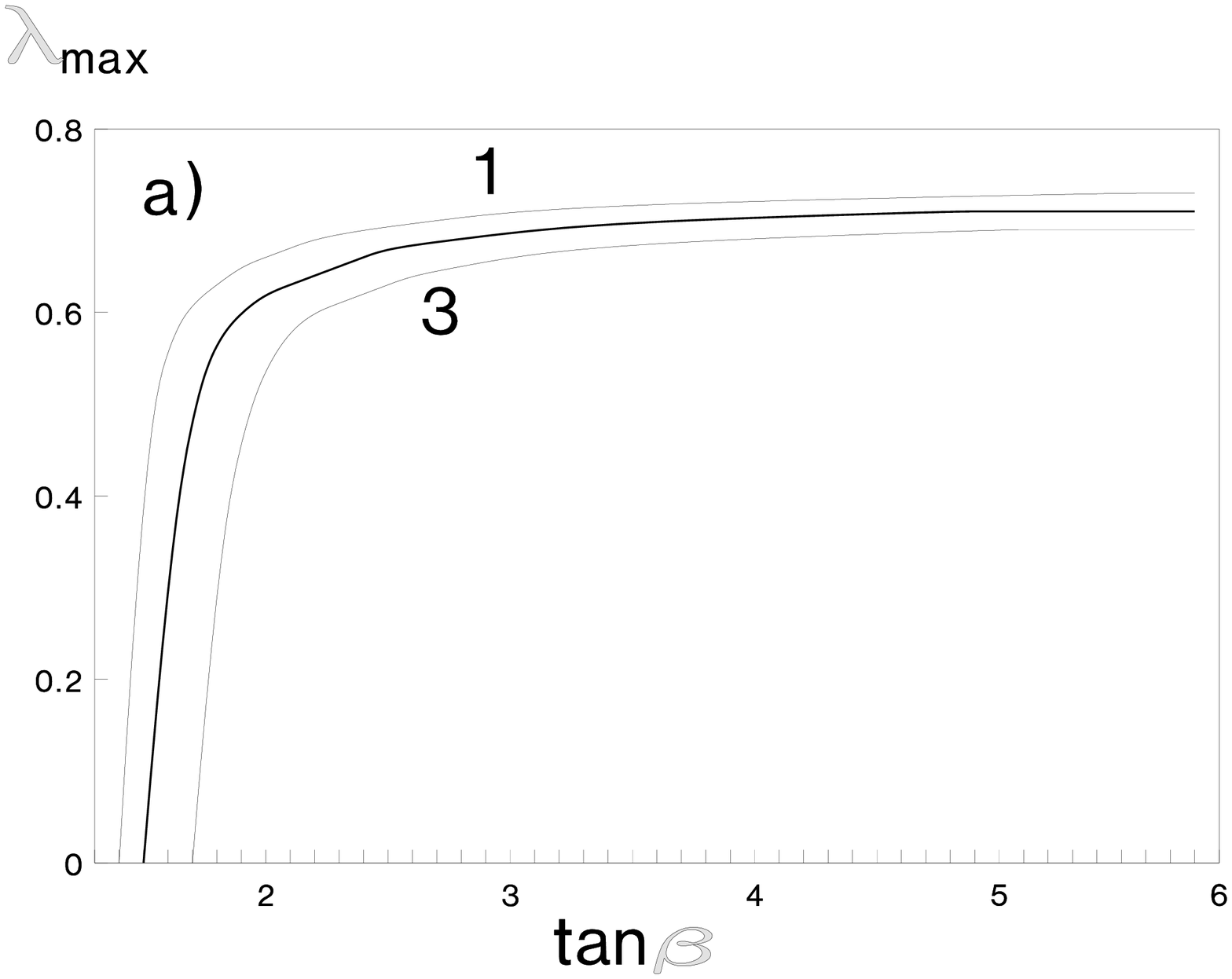}}
\epsfxsize=15cm
\epsfysize=9cm
\mbox{\hskip 0.4in \epsfbox{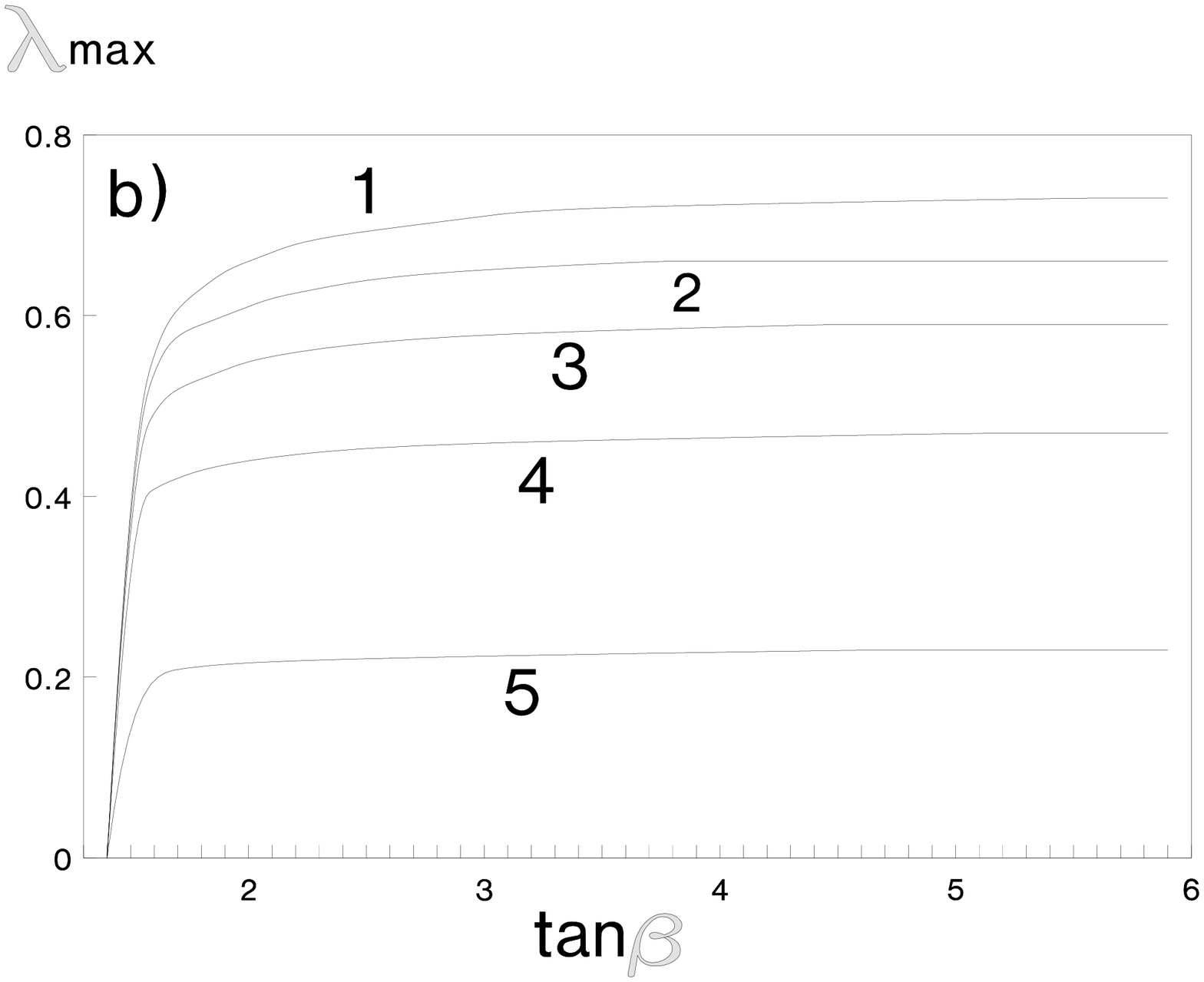}}
\end{figure} 
As one can see from this figure, $|\lambda(m_t)| \leq 0.7$ - this
bound is obtained for $\tan{\beta} \geq 3$ and $\kappa(m_t)=0$. Thus,
the absolute upper boundary on $\lambda$, derived here, is about  20$\%$
lower than IR-quasi fixed point $\lambda=0.87$. Such a difference
arises, when the constraints on $h_t(m_t)$, coming from the recent 
experimental results for the top quark mass, are taken into account. \\
The behaviour of $\lambda_{max}$ with $\tan{\beta}$ is nontrivial. As one
can see from Fig,~1, upper bound on $\lambda$ remains almost unchanged, when
$3 \leq \tan{\beta} \leq 6$, it decreases slowly, when $2 \leq \tan{\beta}
\leq 3$
and, finally, $\lambda_{max}$ is dropped to zero for the lowest allowed
values\footnote{In other words, $\lambda_{max}$ is highly
sensitive to $h_t(m_t)$, when the latter is close to its upper boundary.
On the other hand, the RGE's analysis approach is not
able to make predictions for the couplings with such an accuracy.
This makes
the predictions for $\lambda$ unreliable, when the lowest values of 
$\tan{\beta}$ are considered.
Notice however that our main results are obtained for 
$\tan{\beta} \approx 2$, where such a problem does not exist.} of
$\tan{\beta}$.   \\
It is interesting to investigate also, how the restrictions on $\lambda$ are
affected by the experimental error in the top mass. For this purpose I find
$\lambda_{max}$ for $m_t=160$GeV, $m_t=165$GeV and $m_t=170$GeV. As one can
see from Fig,~1a), either $\lambda_{max}(\tan{\beta})$ or
$\tan{\beta}(\lambda_{max})$ vary only weakly, when the top mass varies form
160GeV to 170GeV. Notice however that this weak variation of
$\lambda_{max}$ (or $\tan{\beta}$) becomes nonnegligible, when being
combined with the sensitivity of the lightest Higgs mass to the top mass
via the radiative corrections \cite{45,46}.  \\
It is clear that the difference between the pole
and the on-shell top masses is nonnegligible too. This is in spite of this
difference affecting only weakly on $\lambda_{max}$: one obtains
$|\lambda(m_t)| \leq 0.65$, when taking $m_t=174$GeV (or
$h_t(m_t) \geq 1/\sin{\beta}$). \\
The restrictions on $\lambda$, discussed above, were derived in the limit
$\kappa=0$. Notice however that they may be considered approximately valid for
$|\kappa(m_t)| < 0.3$. Indeed, as one can see from Fig,~1b), the difference
between the results, derived for $\kappa(m_t)=0$ and $|\kappa(m_t)| = 0.3$ is
negligible for $\tan{\beta}<2$ and only weak for $\tan{\beta} > 2$. More
precisely, for $|\kappa(m_t)|=0.3$ one derives $|\lambda(m_t)| \leq 0.65$
instead of $|\lambda(m_t)| \leq 0.7$. For fixed values of the top mass and
$\tan{\beta}$ such a small difference does not affect the predictions
for the lightest Higgs mass. This allows one to avoid
the problems, connected with the correlations between the predictions for
$\lambda$ and $\kappa$. \\
\\
6. Let us return to the Higgs bosons masses problem in the MSSM and its
simplest extension with the additional Higgs singlet. As it was discussed
previously, the restrictions on the lightest Higgs mass depend on the gauge
coupling constants, on
the Higgs vev's ratio $\tan{\beta}$, on the top quark mass, on the
stops masses, which are
identified here with the supersymmetry breaking scale, on the parameter
$X_t$, given by eq, (3), and (in the NMSSM) on the coupling constant
$\lambda$.
Here I take $1.4 \leq \ \tan{\beta} \leq 6$ (the lower bound
on $\tan{\beta}$ is
derived, using the upper bound on $h_t(m_t)$), $M_{SUSY}=1$TeV,
$X_t=0$ and $X_t=6$. The values of the gauge couplings and the top quark
mass, as well as the restrictions on $\lambda$ were presented in previous
section.
The derived upper bound on the
lightest Higgs mass ($m_{h_{max}}$) as a function of $\tan{\beta}$ is
presented in Fig,~2. \\
\begin{figure}
\caption{Upper bound on the lightest Higgs mass in the MSSM and the NMSSM
a) for $X_t=0$ (no-mixing) b) for $X_t=6$ (maximal mixing). The results
are derived for $m_t=160GeV$ (dashed line), $m_t=165$GeV (solid line) and
for $m_t=170$GeV (dotted line).}
\epsfxsize=15cm
\epsfysize=9cm
\mbox{\hskip 0.4in \epsfbox{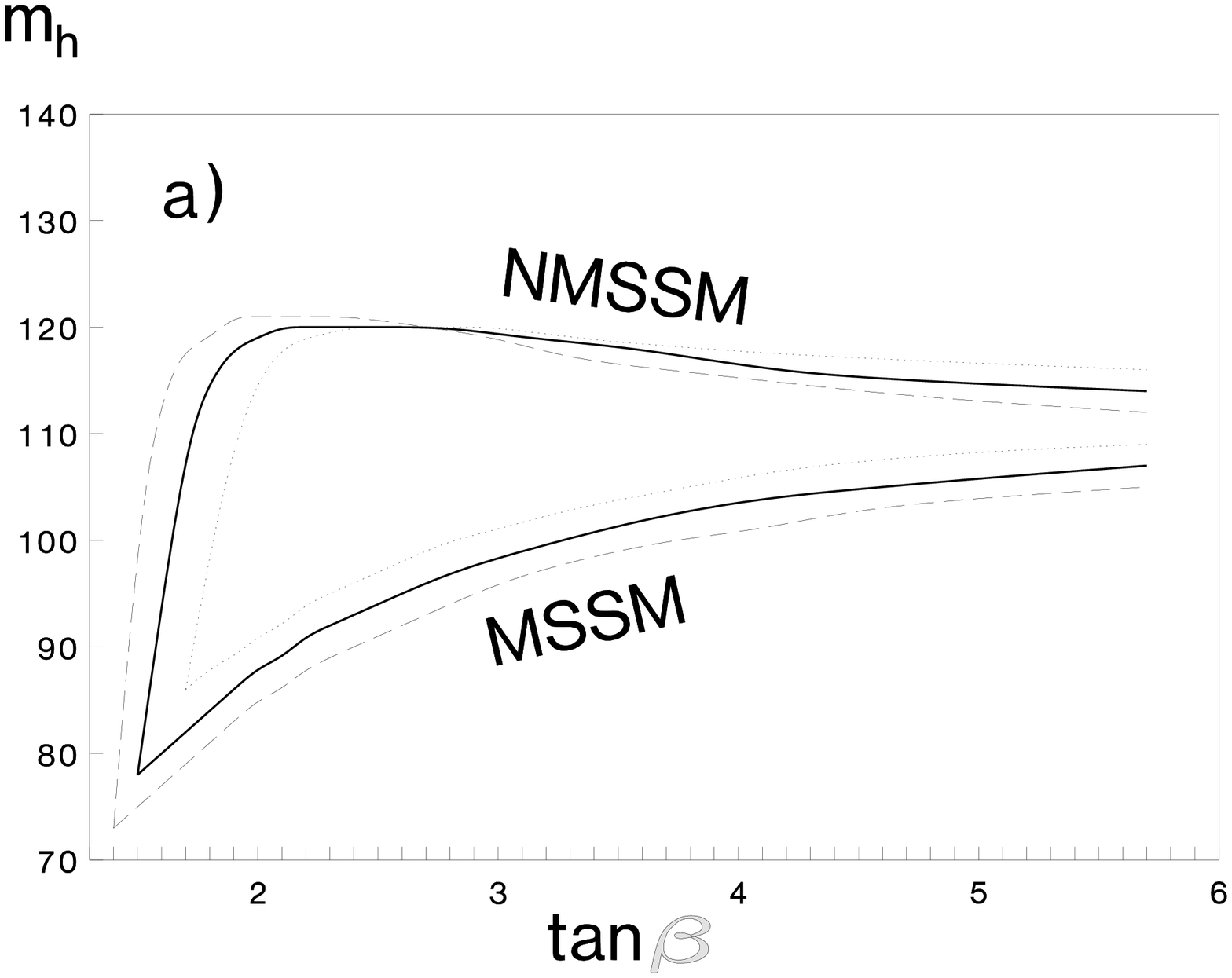}}
\epsfxsize=15cm
\epsfysize=9cm
\mbox{\hskip 0.4in \epsfbox{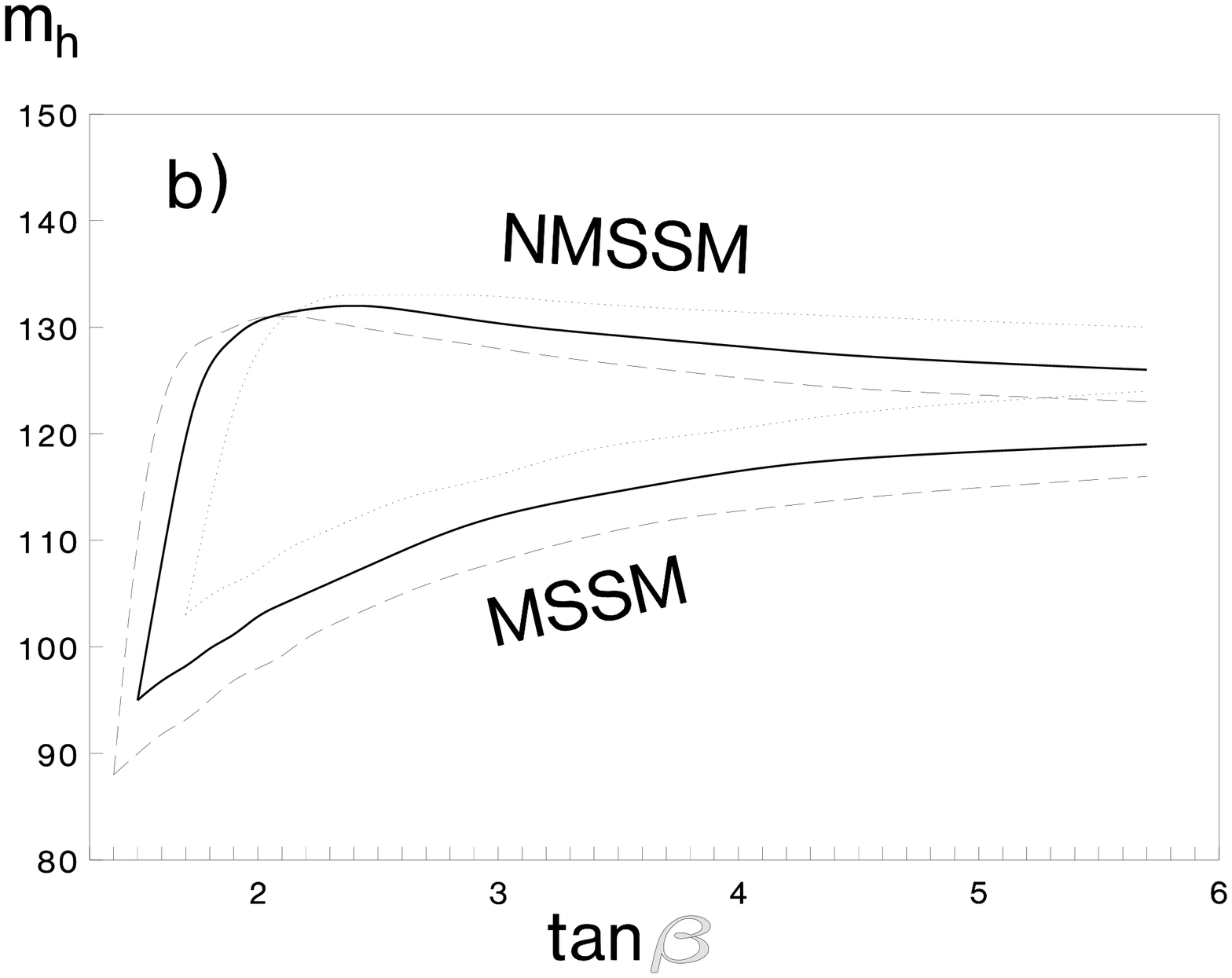}}
\end{figure}
As one can see from this figure, the MSSM and the
NMSSM upper bounds on the lightest Higgs mass are different in general.
This difference  ($\Delta_h$)
is maximal for $\tan{\beta}$ being close to 2. For
lower values of $\tan{\beta}$ the difference between the models predictions
decreases fastly to zero (together with $\lambda_{max}$). For larger values
of $\tan{\beta}$ $\Delta_h$ decreases too:  as one can see from Fig,~2,
the predictions of the models tend slowly to coincide for $\tan{\beta}
\gg 1$.  \\
The results, presented in Fig,~2, are obtained for $m_t=160$GeV,
$m_t=165$GeV and $m_t=170$GeV, for the cases of no-mixing (Fig,~2a) and
maximal mixing (Fig,~2b). The largest difference between the models
predictions is derived for $m_t=160$GeV. Thus, in the case of no-mixing
one obtains $\Delta_h \approx 40$GeV. For $m_t=165$GeV and $m_t=170$GeV
it is obtained $\Delta_h \approx 32GeV$ and $\Delta_h \approx 25$GeV
respectively.
The situation is similar for the case of maximal mixing. Here one obtains
$\Delta_h \approx 35$GeV, 28GeV and 22GeV respectively for $m_t=160$GeV,
165GeV and 170GeV. \\
Thus, the difference between the models predictions varies about 15GeV, when
the top mass varies from 160GeV to 170GeV. This is caused by different
behaviour of the lightest Higgs mass upper bound
with the top mass in the MSSM and in the NMSSM. In the
MSSM the experimental error $\pm 5$GeV in the top mass leads to approximately
the same error in $m_{h_{max}}$. On the contrary, in the NMSSM for
low values of $\tan{\beta}$ $m_{h_{max}}$ either remains unchanged or
decreases with the increasing of the top mass. \\
There is also the difference in the behaviour of the models predictions with
$\tan{\beta}$. As one can see from Fig,~2, in the MSSM $m_{h_{max}}$
increases monotonously with $\tan{\beta}$, so that the absolute upper bound
on $m_h$ is reached for $\tan{\beta} \gg 1$. On the contrary, in the NMSSM
the absolute upper bound is saturated for $\tan{\beta}= 2-2.5$. Such a
difference in the models predictions is connected, in particular, with the
difference between the pole and the on-shell top masses, as well as with
two-loop
order radiative corrections to the lightest Higgs mass. On the absence of
these two effects the NMSSM absolute upper bound on $m_h$ is saturated for
$\tan{\beta} \gg 1$ too \cite{5}. \\
It is interesting to compare the results, derived here in the case of the
NMSSM, to those, reported in literature previously. Let us start with the
case of
no-mixing - in this case it is obtained that $m_h < 120$GeV. The
study of this case, using the RGE's for the Higgs potential effective
coupling constants,
has been carried out in ref. \cite{10} - there the result $m_h < 145$GeV
had been obtained. As one can see, the difference between the results is
25GeV. This difference is caused both by stronger restrictions on
$\lambda$ and by next-to-leading order corrections to the lightest Higgs
mass. These two effects modify the predictions for $m_{h_{max}}$ by
15GeV and 10GeV respectively. \\
In the case of maximal mixing it is obtained $m_h < (130-135)$GeV (depending
on the top mass). This result is compared to those, reported in ref's
\cite{29,11,17} - there the upper bound $m_h < 155$GeV is obtained.
Again, the difference between the results is (20-25)GeV. However,
unlike the case of no-mixing, the dominant contribution
to this difference (15GeV) comes now
from two-loop order radiative corrections. \\
\\
7. Thus, the problem of the restrictions on the lightest Higgs boson mass
in the next-to-minimal supersymmetric standard model has been
analyzed. To obtain these restrictions, the constraints on coupling
constants of theory were taken into account. These constraints come from
the renormalization group equations analysis and from the recent
experimental results for the top mass. The
experimental constraints on the top mass are very important: they make
the restrictions on coupling constants much stronger. The another important
effect is connected with two-loop order radiative corrections to the lightest
Higgs mass. Due to these two effects the derived
NMSSM upper bound on the lightest Higgs mass is about (20-25)GeV lower than
the one, reported in literature previously. \\
There is a large difference between the MSSM and the NMSSM constraints on
the lightest Higgs mass. Namely, the upper bound on the
lightest Higgs mass in the NMSSM can be about 40GeV larger
than in the MSSM. Such a large
difference occurs for $\tan{\beta}$ being close to 2. Due to this difference
the NMSSM with the low $\tan{\beta}$ scenario is, generally speaking, far out
from being excluded.
Notice however that the NMSSM upper bound
on the lightest Higgs mass is saturated in the narrow region of parameters
space, where the singlet Higgs sector is decoupled and the detectable
Higgs bosons masses are described by the MSSM-type mass matrices.  To
compare the models predictions in wider region of parameters space, the
predictions for the supersymmetry breaking parameters of theories are
necessary. \\
\\
\\
\\
\centerline{\bf{Acknowledgements}}
\vspace{0.1cm}
Author is grateful to H. M. Asatrian for stimulating discussions.
Some of results, presented in this paper, are based on the observations,
which I did during the collaboration with F. Schrempp -
I am grateful to him for discussion.
The work was supported by INTAS under Contract INTAS-96-155. \\
\\
\\
\renewcommand{\theequation}{A.\arabic{equation}}
\setcounter{equation}0
\appendix
\begin{center}
    APPENDIX
\end{center}
\vspace{0.1cm}
At tree level the  $3 \times 3$ symmetric mass matrix for the fields
$\Phi_1$, $\Phi_2$ and $N_1$ is the following:
\begin{eqnarray}
\nonumber
M_{S_{11}}^2 &=& M_Z^2 \cos^2 2\beta + \lambda^2 \eta^2 \sin^2 2\beta \\
\nonumber
M_{S_{12}}^2 &=& -(M_Z^2-\lambda^2 \eta^2) \sin2\beta \cos2\beta\\
\nonumber
M_{S_{13}}^2 &=& 2 \lambda^2 v_3 \eta + (A_\lambda + 2 \kappa v_3) \lambda
\eta \sin2\beta \\
M_{S_{22}}^2 &=& (M_Z^2-\lambda^2 \eta^2) \sin^2 2\beta -2 \lambda v_3
\frac{A_\lambda+\kappa v_3}{\sin2\beta}\\
\nonumber
M_{S_{23}}^2 &=& -(A_\lambda + 2 \kappa v_3) \lambda    
\eta \cos2\beta \\
\nonumber
M_{S_{33}}^2 &=& 4 \kappa^2 v_3^2 + \kappa v_3 A_\kappa - \frac{A_\lambda
\lambda
\eta^2 \sin2\beta}{2 v_3}
\end{eqnarray}
where $A_\lambda$, $A_\kappa$ are trilinear SUSY breaking parameters of the
Higgs potential \cite{9,10,11,15}, $v_3$ is the singlet vev. The fields
$\Phi_1$, $\Phi_2$, $N_1$ are given by equation (4).

\end{document}